\documentclass[10pt, pra, aps, twocolumn, showpacs, floatfix]{revtex4-1}
\usepackage{bbm}
\usepackage{epsfig}
\usepackage{epstopdf}
\usepackage{graphicx, subfigure}
\usepackage{amsmath,amssymb}
\usepackage{color}
\usepackage{hyperref}
\usepackage{appendix}
\usepackage{multirow}
\graphicspath{{images/}}
\DeclareMathOperator{\sech}{sech}

\begin{document}

\title{Stroboscopically Robust Gates For Capacitively Coupled Singlet-Triplet Qubits}

\author{R.~K.~L.~Colmenar}
\email{ralphkc1@umbc.edu}
\author{J.~P.~Kestner}
\affiliation{Department of Physics, University of Maryland Baltimore County, Baltimore, MD 21250, USA}

\begin{abstract}
Recent work on Ising-coupled double-quantum-dot spin qubits in GaAs with voltage-controlled exchange interaction
has shown improved two-qubit gate fidelities from the application of oscillating exchange along with a strong
magnetic field gradient between adjacent dots~\cite{Nichol_2017}. By examining how noise propagates in the time-evolution operator
of the system, we find an optimal set of parameters that provide passive stroboscopic circumvention of errors in
two-qubit gates to first order. We predict over 99\% two-qubit gate fidelities in the presence of quasistatic and
1/\textit{f} noise, which is an order of magnitude improvement over the typical unoptimized
implementation.

\end{abstract}

\maketitle

\section{Introduction}
\label{sec:Intro}
Quantum dot spin qubits provide a promising platform for quantum computing due to their potential scalability and
relatively long coherence times. For single-spin qubits \cite{Loss1998}, one-qubit operations with gate fidelities exceeding the fault-tolerant
threshold have been realized in single-spin qubits \cite{Veldhorst_2014}, but two-qubit gates have much lower fidelities
\cite{Watson_2018,Zajac_2018}.  Likewise, for singlet-triplet
spin qubits~\cite{Petta_2005,Maune_2012}, which we focus on below, a recent two-qubit experiment reported only up to 90\% entangling gate fidelity \cite{Nichol_2017}.
This can be improved by circumventing the effects of the two main noise sources, namely fluctuations in the electric confining
potential and fluctuations in the Zeeman energy difference between the quantum dots.

The fluctuation in the confining potential is often
attributed to thermal fluctuations in the occupation of nearby charge traps, i.e., charge noise, thus leading to
fluctuations in the local electric field~\cite{Kuhlmann_2013}. Relative to the time-scale of spin qubit
rotation times, these fluctuations can be treated quasistatically as a first approximation, but the actual power spectral density of charge noise in these qubit systems
has been measured to behave like $1/f^{0.7}$ in GaAs~\cite{Dial_2013} and $1/f$ in Si~\cite{Eng_2015,Yoneda_2018} out to tens or even hundreds of kHz. The quasistatic part of
the noise can be addressed by applying composite pulse sequences, where noisy gate operations are applied sequentially such that the gate errors conspire to cancel one another.
These sequences, however, typically only suppress noise that is slow on the timescale of the sequence, and amplify noise that is faster~\cite{Green_2013}.

The Zeeman fluctuations manifest in two ways depending on how the gradient is generated.  When the gradient comes from the
Overhauser effect due to the hyperfine coupling of the dot electron with the nuclear spin of the host semiconductor,
such as in GaAs-based architectures using dynamical nuclear spin polarization \cite{Reilly_2008,Cywinski_2009,Barnes_2012}, electron-mediated nuclear spin flip-flops produce
$1/f^2$ noise \cite{Medford_2012,Malinowski_2017} that is essentially quasistatic.  When the gradient comes from a micromagnet structure \cite{Yoneda_2015}, as used in some GaAs devices \cite{Ladriere2008,Brunner_2011} and which is necessary for silicon-based architectures with far fewer spinful nuclei \cite{Wu2014}, it is possible for charge noise to also couple in via small shifts in the dot position, again resulting in higher
frequency noise~\cite{Kawakami_2016}.

Two-qubit gate fidelity in singlet-triplet systems is mostly limited by charge noise when the qubit dynamics is dominated by the exchange interaction \cite{Petta_2005,Shulman_2012}.
Recent work on capacitively-coupled, double-quantum dot spin qubits with gate-controlled exchange coupling between the spins has demonstrated suppression of charge noise by
applying a strong magnetic gradient between the two dots in each qubit that is much stronger than the exchange interaction~\cite{Nichol_2017}. An analytical
expression for the full time-evolution operator of this particular system can be obtained by using the rotating-wave approximation (RWA)~\cite{Calderon_2017}.

In this work, we analyze how perturbations in the control parameters of a capacitively-coupled singlet-triplet system affect the time-evolution and present a strategy to minimize those effects. In Sec.~\ref{sec:Evolution},
we derive the time-evolution operator using the RWA. We consider in Sec.~\ref{sec:Error_Channels} two different parameter regimes for two qubits with similar energy
splitting: when the magnetic field gradient dominates the splitting, and when the exchange interaction dominates instead. We calculate the leading order errors and show
that certain parameter choices result in a synchronization of oscillating error terms such that a passive reduction of gate errors occurs at specific times. In Sec.
\ref{sec:Simulations} we examine the effects of the optimization in the presence of both quasistatic noise and 1/\textit{f} noise. We find that our optimization
isolates the effects of noise into particular SU(4) basis elements, allowing us to prescribe composite pulse sequences to mitigate the remaining errors. In principle, this work allows the improvement of experimental two-qubit gate fidelities to above 99\%. While most of our work is presented in the limit of zero pulse rise time, we show in App.~\ref{app:ramp} that typical finite
rise times do not pose a challenge to the stroboscopic error suppression.

\section{The Time-Evolution Operator}
\label{sec:Evolution}
We consider a system of capacitively-coupled singlet-triplet qubits, which corresponds directly to the experimental setup in Ref.~\cite{Nichol_2017}, but our results are also applicable
to any system similarly described by a static Ising coupling and local driving fields. The effective two-qubit Hamiltonian is given by
\begin{equation}
\label{eq:Hamiltonian}
\mathcal{H} = \sum_{i=1}^{2} \bigg( \frac{J_{i} + j_{i}\cos[\omega_{i}t]}{2}\sigma_{Z}^{(i)} +
\frac{h_{i}}{2}\sigma_{X}^{(i)}\bigg) + \alpha\,\sigma_{ZZ},
\end{equation}
where $\sigma_{ij} \equiv \sigma_{i}^{(1)} \otimes \sigma_{j}^{(2)}$ with $\{i,j\} \in \{I,X,Y,Z\}$ collectively form a 15-dimensional SU(4) basis. The exchange interaction between
two spins in the $i^{\text{th}}$ qubit is a function of the difference in electrochemical potential between the dots, $\varepsilon_i$, which can vary in time. By
oscillating $\varepsilon_i$, the exchange is caused to oscillate at a driving frequency $\omega_i$, which makes the effective exchange interaction oscillate about an
average value $J_i$ with an amplitude $j_i$. The static, longitudinal magnetic field gradient is denoted by $h_i$; this can be generated by using either a micromagnet
or, in GaAs, through the hyperfine interaction between the dot electrons and the nuclear spins in the semiconductor. Thus, the static part of a qubit's total energy
splitting is $\Omega_i \equiv \sqrt{h_{i}^{2} + J_{i}^{2}}$. Finally, $\alpha$ is the electrostatic coupling strength between the adjacent qubits, which is proportional
to the product of the two qubits' electric dipole moments.

Ref.~\cite{Calderon_2017} reported an approximate time-evolution operator for the aforementioned Hamiltonian using the RWA. There it was implicitly assumed that
$\frac{j_i J_i}{2\Omega_i} \ll \Omega_i$. We lift this assumption and apply the same formalism to find a more general description of the time evolution. We begin by
first performing a local rotation to align the x-axis along the vector sum of the combined local static fields
\begin{equation}
\label{eq:first_rot}
U = \exp\left[\frac{\imath}{2}\sum_{i=1}^{2} \phi_{i} \sigma_{Y}^{(i)} \right] U_1 \exp\left[-\frac{\imath}{2}\sum_{i=1}^{2} \phi_{i} \sigma_{Y}^{(i)} \right],
\end{equation}
where $\phi_{i} \equiv 	\tan^{-1}(J_{i}/h_{i})$ and $U$ is the lab-frame propagator. We then transform to the rotating frame
\begin{equation}
\label{eq:second_rot}
U_1 = \exp \left[ -\imath \sum_{i=1}^{2} \left( \frac{\omega_i t + \xi_i(t)}{2} \right) \sigma_{X}^{(i)} \right] U_2,
\end{equation}
where the inclusion of $\xi_i(t) = \frac{j_i J_i \sin \left( \omega_i t\right)}{\omega_i \Omega_i}$ generalizes Ref.~\cite{Calderon_2017}.
We perform the RWA by doing a coarse-grain time-average over a time scale $1/\alpha \gg \tau \gg \textrm{max}\{1/\omega_i\}$. The addition of $\frac{\xi_i(t)}{2}$
in the local rotation causes some of the terms in the rotating-frame Hamiltonian to have nontrivial averages. The time-averaged propagator is given by
\begin{multline}
\label{eq:first_RWA}
U_2 = \exp \Bigg[ -\imath t  \Bigg( \sum_{i=1}^{2} \left(\chi_i \sigma_{Z}^{(i)} + \frac{\Omega_i -\omega_i}{2} \sigma_{X}^{(i)}\right) \\
- \frac{h_1 J_2 \alpha}{\Omega_1 \Omega_2} \mathcal{J}_{1}\left[\frac{j_1 J_1}{\omega_1 \Omega_1}\right] \sigma_{ZX}
- \frac{h_2 J_1 \alpha}{\Omega_1 \Omega_2} \mathcal{J}_{1}\left[\frac{j_2 J_2}{\omega_2 \Omega_2}\right] \sigma_{XZ} \\
+ \frac{J_1 J_2 \alpha}{\Omega_1 \Omega_2} \sigma_{XX}
+ \frac{h_1 h_2 \alpha}{2\Omega_1 \Omega_2} \left( \mathcal{I}_{YY}\sigma_{YY}+\mathcal{I}_{ZZ}\sigma_{ZZ} \right) \Bigg)\Bigg],
\end{multline}
where $\mathcal{J}_{i}[z]$ is the $\mathit{i^{th}}$ order Bessel function of the first kind,
$\chi_i \equiv \frac{h_i \omega_i \mathcal{J}_{1}\left[\frac{j_i J_i}{\omega_i \Omega_i}\right]}{2 J_i}$ is the Rabi frequency, and
\begin{align}
\label{eq:integralYY}
\mathcal{I}_{YY} &= \frac{1}{\tau}\int_0^{\tau} 2\sin \left( \omega_1 t + \xi_1 \right)
	 \sin\left( \omega_2 t + \xi_2 \right) \mathrm{d}t \\
\label{eq:integralZZ}
\mathcal{I}_{ZZ} &= \frac{1}{\tau}\int_0^{\tau} 2\cos \left( \omega_1 t + \xi_1 \right)
	 \cos\left( \omega_2 t + \xi_2 \right) \mathrm{d}t.
\end{align}
We require $\omega_i \gg \left\lbrace\left|\frac{h_i j_i}{2\Omega_i}\right|, \alpha \right\rbrace$ to ensure the validity of the RWA.

To gain a better understanding of the entangling dynamics, we take another transformation to eliminate the remaining local operators in the Hamiltonian:
\begin{equation}
\label{eq:third_rot}
U_2 = \exp\left[ -\imath t \sum_{i=1}^{2} \left(\frac{\Omega_i - \omega_i}{2} \sigma_{X}^{(i)} +  \chi_i \sigma_{Z}^{(i)} \right) \right] U_3.
\end{equation}
We set the control field at resonance with the energy splitting, $\omega_i = \Omega_i$, thus eliminating the $\sigma_{X}^{(i)}$ terms. Note that by completely dropping
this off-resonant term below, we have limited the validity of our analysis to cases where perturbations in $\Omega_i$ are much less than $\chi_i$.  Lifting this assumption
would not permit us to obtain a time-independent Hamiltonian. Nonetheless, this is not an unrealistic assumption.  At this point, we can proceed the same way as in
Ref.~\cite{Calderon_2017}. We apply another round of the RWA which requires $|\chi_i| \gg \alpha$. If $||\chi_1| - |\chi_2|| \ll \alpha$, the average time-evolution
operator is given by
\begin{equation}
\begin{aligned}
\label{eq:second_RWA_v1}
U_3 &= \exp \bigg[\frac{-\imath\alpha t}{2}\bigg( \frac{h_1 h_2 \mathcal{I}_{YY} + 2J_1 J_2}{2\Omega_1 \Omega_2} (\sigma_{XX}+\sigma_{YY})\\
	&+ \frac{h_1 h_2}{\Omega_1 \Omega_2}\mathcal{I}_{ZZ} \sigma_{ZZ}\bigg)\bigg],
\end{aligned}
\end{equation}
but if $||\chi_1| - |\chi_2|| \gg \alpha$, we instead have
\begin{equation}
\label{eq:second_RWA_v2}
U_3 = \exp \bigg[-\imath t\frac{\alpha h_1 h_2}{2\Omega_1 \Omega_2}\mathcal{I}_{ZZ} \sigma_{ZZ}\bigg],
\end{equation}
This reduces to the result of Ref.~\cite{Calderon_2017} in the regime $h_i \gg J_i$, which is experimentally relevant \cite{Nichol_2017}, but it becomes quite
different when the exchange is dominant, as in earlier experiments \cite{Petta_2005,Shulman_2012}.

The entangling dynamics depend on whether the qubit energy splittings, $\Omega_i$, are nearly equal or not. If the difference between the two energy splittings
is much larger than $\alpha$, $|\Omega_1 - \Omega_2| \gg \alpha$, $\mathcal{I}_{ZZ}$ and $\mathcal{I}_{YY}$ become small. Looking at Eqs.~\eqref{eq:second_RWA_v1}
and \eqref{eq:second_RWA_v2}, one can avoid a suppressed coupling rate by setting the Rabi frequencies equal to one another, $\chi_1=\chi_2$, and operating in
the large exchange regime, $J_i \gg h_i$. On the other hand, if the two qubits have similar energy splittings, the effective coupling rate is $\sim\alpha$ regardless
of which parameter dominates.

\section{First-Order Error Channels}
\label{sec:Error_Channels}

As previously mentioned, the magnetic field gradient, $h_i$, in singlet-triplet systems is produced by either micromagnets, as demonstrated in a silicon-based experiment \cite{Wu2014},
or the hyperfine interaction between the quantum dot electron and the nuclear spins, as has often been used in the case of GaAs \cite{Reilly_2008,Cywinski_2009,Barnes_2012}. Whereas the latter case allows some fine-tuning of $h_i$
through dynamic nuclear polarization, the same is not true for micromagnets. Thus, we consider two main cases of experimental relevance -- when $h_i$ is tunable
and when it is not. Furthermore, the sensitivity of the qubits to fluctuations depends on the parameter regime at work. If $J_i$ and $h_i$ are completely uncorrelated,
the fluctuation on the qubit energy splitting is given by
\begin{equation}
\delta\Omega_i^2 = \frac{J_i^2 \delta J_i^2 + h_i^2 \delta h_i^2}{\Omega_i^2}.
\end{equation}
Note that when either $J_i$ or $h_i$ completely dominates the energy splitting, the noise due to the weaker one is suppressed by a factor of their ratio. We know
from experiments that $\delta h_i$ is mostly quasistatic on the timescale of the gates~\cite{Medford_2012,Malinowski_2017} and $\delta J_i$ contains both a quasistatic and a
1/\textit{f} component~\cite{Dial_2013}. Thus, it is best to suppress the 1/\textit{f} $\delta J_i$ errors by choosing $h_i \gg J_i$ and then correct the residual quasistatic errors with
spin echo protocols. This is consistent with the improvement reported in Ref.~\cite{Nichol_2017} when the magnetic field gradient was increased.

As discussed in the previous section, rapid entanglement in the $h_i \gg J_i$ regime only occurs when the two qubit energy splittings are tuned close to one
another ($h_1 \approx h_2$). If one is forced to work with fixed but very different gradients ($|h_1 - h_2| \gtrsim \min\{h_i\}$), which is a possible scenario when
micromagnets are used, then one must work in the $J_i \gg h_i$ regime. Therefore, we will limit our discussion to these two cases: when $h_i$ is dominant and when
$J_i$ is dominant. We assume similar qubit energy splittings in both cases for convenience, particularly when simplifying Eqs.~\eqref{eq:integralYY} and \eqref{eq:integralZZ}.

\subsection{Similar qubits with $\mathbf{h_i} \gg \mathbf{J_i}$}
\label{subsec:similar}

\begin{table*}
\caption{First-order errors obtained by projecting $\Delta$ onto an SU(4) basis formed by Kronecker products of Pauli operators.}
\label{table:TotError}
\begin{center}
\begin{tabular}{c c}
	\hline
	\hline
	$\sigma_{IX}$ & $\left( \frac{\imath(h_{2}\delta J_{2} - J_{2} \delta h_{2})\cos(\omega t)}{2\Omega_{2}^{2}}
			- \frac{\imath (h_2 \delta h_2 + J_2 \delta J_2)}{h_2 j_2}\right) \cos\big(\frac{h_1 h_2 \alpha t}
			{\Omega_1 \Omega_2}\big)\sin(2\chi_2 t) $ \\
	$\sigma_{IY}$ & $\frac{\imath(h_{2}\delta J_{2} - J_{2} \delta h_{2})\Big(\cos\big(\frac{h_1 h_2 \alpha t}
			{\Omega_1 \Omega_2}\big) \cos(\omega t) \cos(2\chi_2 t)-1\Big)}{2\Omega_{2}^{2}} +\frac{2\imath
			(h_2 \delta h_2 + J_2 \delta J_2)\cos\big(\frac{h_1 h_2 \alpha t}{\Omega_1 \Omega_2}\big)\sin^{2}
			(\chi_2 t)}{h_2 j_2}$ \\
	$\sigma_{IZ}$ & $\frac{\imath(h_{2}\delta J_{2} - J_{2} \delta h_{2}) \big(j_{2}J_{2}t -
			2\Omega_2\sin(\omega t)\big)}{4\Omega_{2}^{3}} - \frac{\imath h_2 t \delta j_2}
			{4\Omega_2}$\\
	$\sigma_{XI}$ & $\left( \frac{\imath(h_{1}\delta J_{1} - J_{1} \delta h_{1})\cos(\omega t)}{2\Omega_{1}^{2}}
			- \frac{\imath(h_1 \delta h_1 + J_1 \delta J_1)}{h_1 j_1}\right) \cos\big(\frac{h_{1}h_{2}\alpha t}
			{\Omega_{1}\Omega_{2}}\big) \sin(2\chi_{1}t)$\\
	$\sigma_{XX}$ & 0\\
	$\sigma_{XY}$ & 0\\
	$\sigma_{XZ}$ & $\left(\frac{\imath (h_{1} \delta J_{1}-J_{1}\delta h_{1})\cos(\omega t)\cos(2\chi_{1}t)}{2\Omega_{1}^{2}}
			+\frac{2\imath (h_1 \delta h_1 + J_1 \delta J_1)\sin^{2}(\chi_1 t)}{h_1 j_1}\right) \sin\big(\frac{h_{1}h_{2}\alpha t}
			{\Omega_{1}\Omega_{2}}\big)$\\
	$\sigma_{YI}$ & $\frac{\imath(h_{1}\delta J_{1} - J_{1} \delta h_{1})\Big(\cos\big(\frac{h_1 h_2 \alpha t}
	{\Omega_1 \Omega_2}\big) \cos(\omega t) \cos(2\chi_1 t)-1\Big)}{2\Omega_{1}^{2}} + \frac{2\imath
			(h_1 \delta h_1 + J_1 \delta J_1)\cos\big(\frac{h_1 h_2 \alpha t}{\Omega_1 \Omega_2}\big)\sin^{2}
			(\chi_1 t)}{h_1 j_1}$ \\
	$\sigma_{YX}$ & 0\\
	$\sigma_{YY}$ & 0\\
	$\sigma_{YZ}$ & $\left( \frac{\imath (J_{1}\delta h_{1} - h_{1} \delta J_{1}) \cos(\omega t)}{2\Omega_{1}^{2}}
			+ \frac{\imath(h_1 \delta h_1 + J_1 \delta J_1)}{h_1 j_1}\right) \sin\big(\frac{h_{1}h_{2}\alpha t}
			{\Omega_{1}\Omega_{2}}\big) \sin(2\chi_{1}t)$\\
	$\sigma_{ZI}$ & $\frac{\imath(h_{1}\delta J_{1} - J_{1} \delta h_{1}) \big(j_{1}J_{1}t-2\Omega_{1}\sin
			(\omega t)\big)}{4\Omega_{1}^{3}} - \frac{\imath h_1 t \delta j_1}{4\Omega_1}$\\
	$\sigma_{ZX}$ & $\left(\frac{\imath (h_{2} \delta J_{2}-J_{2}\delta h_{2})\cos(\omega t)\cos(2\chi_{2}t)}{2\Omega_{2}^{2}}
			+ \frac{2\imath(h_2 \delta h_2 + J_2 \delta J_2)\sin^{2}(\chi_2 t)}{h_2 j_2}\right) \sin\big(\frac{h_{1}h_{2}\alpha t}
			{\Omega_{1}\Omega_{2}}\big)$\\
	$\sigma_{ZY}$ & $\left(\frac{\imath (J_{2}\delta h_{2} - h_{2} \delta J_{2})\cos(\omega t)\sin(2\chi_{2}t)}{2\Omega_{2}^{2}}
			+ \frac{\imath(h_2 \delta h_2 + J_2 \delta J_2)\sin(2\chi_2 t)}{h_2 j_2}\right)  \sin\big(\frac{h_{1}h_{2}\alpha t}
			{\Omega_{1}\Omega_{2}}\big)\sin(2\chi_{2}t)$\\
	$\sigma_{ZZ}$ & $\frac{\imath h_2 J_1 \alpha t (h_1 \delta J_1 - J_1 \delta h_1)}{2\Omega_{1}^{3} \Omega_{2}}
			+ \frac{\imath h_1 J_2 \alpha t (h_2 \delta J_2 - J_2 \delta h_2)}{2\Omega_{1}\Omega_{2}^{3}} -
			\frac{\imath h_1 h_2 t \delta\alpha}{2\Omega_1 \Omega_2}$\\
	\hline
\end{tabular}
\end{center}
\end{table*}

We consider a system of similar qubits ($\Omega_1 = \Omega_2$) where the magnetic field gradient dominates the energy splitting ($h_i \gg J_i, \Omega_i \simeq h_i$)
and the driving frequencies are equal and at resonance with the energy splitting ($\omega_1 = \omega_2 \equiv \omega = \Omega_i$) in the absence of noise. For
simplicity, we take the case where the Rabi frequencies of the two qubits are dissimilar (Eq.~\eqref{eq:second_RWA_v2}), although our analysis can be extended
to the similar Rabi case easily. In this parameter regime, we can expand $\mathcal{J}_{1}\left[z\right]$ to first-order and obtain
$\chi_i \approx\frac{h_i j_i}{4\Omega_i}$, and $\xi_i(t) \approx 0$ which allows us to evaluate $\mathcal{I}_{YY} = \mathcal{I}_{ZZ} \approx 1$. Thus, combining
Eqs.~\eqref{eq:first_rot}, \eqref{eq:second_rot}, \eqref{eq:third_rot}, and \eqref{eq:second_RWA_v2}, the total time-evolution can be written as
\begin{equation}
\label{eq:lab-sol}
U(t) = R_{1}(t) \exp\bigg[-\imath t\frac{\alpha h_1 h_2}{2 \Omega_1 \Omega_2}\sigma_{ZZ} \bigg] R_{2}(t),
\end{equation}
where the purely local operators $R_{1}(t)$ and $R_{2}(t)$ are given by
\begin{equation}
\label{eq:locals}
\begin{aligned}
R_{1}(t) = &\exp\Bigg[\frac{\imath}{2}\sum_{i=1}^{2}\phi_{i}\sigma_{Y}^{(i)}\Bigg]
\exp \left[ -\imath \sum_{i=1}^{2} \frac{\omega_i t}{2} \sigma_{X}^{(i)} \right] \\
\times &\exp\left[ -\imath t \sum_{i=1}^{2} \chi_i \sigma_{Z}^{(i)} \right], \\
R_{2}(t) = &\exp\Bigg[\frac{-\imath}{2}\sum_{i=1}^{2}\phi_{i}\sigma_{Y}^{(i)}\Bigg].
\end{aligned}
\end{equation}
Since Eq.~\eqref{eq:lab-sol} is already canonically decomposed into local and nonlocal parts~\cite{Kraus_2001}, it is clear to see how to ``undo" the local part
of the evolution that accompanies the entangling gate. By applying additional local operations, $R_1^{\dagger}$ and $R_2^{\dagger}$, in the absence of coupling,
we obtain a purely nonlocal $\sigma_{ZZ}$ gate,
\begin{equation}
\label{eq:nonlocal}
R_{1}^{\dagger}(t) U(t) R_{2}^{\dagger}(t) = \exp\bigg[-\imath t\frac{\alpha h_1 h_2}{2 \Omega_1 \Omega_2}\sigma_{ZZ} \bigg].
\end{equation}
So far we have been careful to distinguish between $\Omega_1$ and $\Omega_2$ so as to allow for the perturbative effect of noise, but other than that we have not
discussed the effect of such a perturbation.  Noise during the original entangling operation produces errors in both the nonlocal phase of Eq.~\eqref{eq:lab-sol} and in its accompanying
local operations given in Eq.~\eqref{eq:locals}. The pre- and post-applied locals, $R_i^{\dagger}$, only undo the ideal local rotations accompanying the entangling gate,
but any random perturbations are left uncanceled.  By expanding each term in Eqs.~\eqref{eq:lab-sol} and \eqref{eq:locals} to first order in perturbations $\delta J_{i}$, $\delta j_{i}$,
$\delta h_{i}$, and $\delta\alpha$, and commuting all of the perturbations to the right, we may write the effect of the noise in the form
\begin{equation}
\label{eq:nonlocal1}
\begin{aligned}
U_{nl}(t) &= R_{1}^{\dagger}(t) U(t) \left(\mathbbm{1} + \Delta_0\left(t\right)\right) R_{2}^{\dagger}(t) \\
&= \exp\bigg[-\imath t\frac{\alpha h_1 h_2}{2 \Omega_1 \Omega_2}\sigma_{ZZ} \bigg]\left(\mathbbm{1} + \Delta\left(t\right)\right) \\
&\simeq \exp\bigg[-\imath t\frac{\alpha}{2}\sigma_{ZZ} \bigg]\left(\mathbbm{1} + \Delta\left(t\right)\right).
\end{aligned}
\end{equation}
where $\mathbbm{1}$ is the identity operator, $\Delta_0$ contains the first-order perturbation of the physical entangling operation $U$, and
$\Delta \equiv R_2 \Delta_0 R_{2}^{\dagger}$ is the resulting perturbation in the purely nonlocal operation. The approximate equality makes use of the fact that
powers of $J_i/h_i$ are negligibly small compared to the dominant errors we wish to correct.  The error $\Delta$ due to the perturbations is reported in
Table~\ref{table:TotError} in terms of its projections onto the 15 SU(4) basis elements, henceforth referred to as error channels,
\begin{equation}
\Delta = \frac{1}{4}\sum_{ij}\mathbf{tr}(\sigma_{ij} \Delta)\sigma_{ij}.
\end{equation}

One prominent feature of these error channels is their oscillatory behavior. Notice that one can, for example, choose parameters such
that $\sin(\chi_i t) = 0$ at the end of the entangling gate. By doing so, one effectively eliminates several error terms in Table~\ref{table:TotError}. If we also
choose parameters such that $\cos(\omega t)=0$ at the time that the gate is complete, all but five of the error channels in Table~\ref{table:TotError}
($\sigma_{ZI}$, $\sigma_{IZ}$, $\sigma_{YI}$,$\sigma_{IY}$, and $\sigma_{ZZ}$) will be synchronized to vanish at the gate time. We are thus left with a gate that
is partially corrected, for both quasistatic and 1/\textit{f} noise.  This stroboscopic circumvention of error requires no knowledge of the errors involved, only that
they are small enough for the higher-order terms in the error expansion to remain insignificant.

Specifically, stroboscopic error elimination can be achieved by choosing
\begin{align}
t &= (m+1/2) \pi/\omega,\\
j_{i} &=\frac{4 n_{i}\Omega_{i}\omega}{h_{i} (m+1/2)} \simeq \frac{4 n_{i}\omega}{(m+1/2)},
\label{eq:robustness}
\end{align}
where $m$ and $n_i$ are integers. We also want to produce a given nonlocal phase, $\exp[\imath \frac{\theta}{2} \sigma_{ZZ}]$, at the end of the operation. So,
we have another constraint from Eq.~\eqref{eq:nonlocal}, which we can satisfy to good approximation by choosing $m$ such that
\begin{equation}
\label{eq:objective}
\left| \frac{(m+1/2)\pi}{\omega}\alpha - \theta \right|.
\end{equation}
is minimized. Due to the typically weak coupling, $\alpha/\omega \ll 1$, the minimum value is likewise small and occurs at a large value of integer $m$
(corresponding to a gate time containing many cycles of the driving field).

\begin{table*}
\caption{The same errors reported in Table~\ref{table:TotError} after substituting the optimized parameters.}
\label{table:TotError2}
\begin{center}
\begin{tabular}{c c}
	\hline
	\hline
$\sigma_{IY}$ & $\imath \frac{J_2 \delta h_2 - h_2 \delta J_2}{2\Omega_{2}^{2}}$\\
$\sigma_{IZ}$ & $\imath \frac{\big((-1)^{m}h_2 -2n_{2}J_{2}\pi\big)(J_2 \delta h_2 - h_2 \delta J_2)}{2h_{2}\Omega_{2}^{2}} - \frac{\imath h_2 t \delta j_2}
			{4\Omega_2}$\\
$\sigma_{YI}$ & $\imath \frac{J_1 \delta h_1 - h_1 \delta J_1}{2\Omega_{1}^{2}}$\\
$\sigma_{ZI}$ & $\imath \frac{\big((-1)^{m}h_1 -2n_{1}J_{1}\pi\big)(J_1 \delta h_1 - h_1 \delta J_1)}{2h_{1}\Omega_{1}^{2}} - \frac{\imath h_1 t \delta j_1}{4\Omega_1}$\\
$\sigma_{ZZ}$ & $\frac{\imath h_2 J_1 \alpha t (h_1 \delta J_1 - J_1 \delta h_1)}{2\Omega_{1}^{3} \Omega_{2}}
			+ \frac{\imath h_1 J_2 \alpha t (h_2 \delta J_2 - J_2 \delta h_2)}{2\Omega_{1}\Omega_{2}^{3}} -
			\frac{\imath h_1 h_2 t \delta\alpha}{2\Omega_1 \Omega_2}$\\
	\hline
\end{tabular}
\end{center}
\end{table*}

As mentioned earlier, we must take care to stay within a parameter regime where the RWA is valid.  We use some of the remaining free parameters to ensure that the
RWA remains valid for the choices above that lead to error cancellation. We enforce the RWA condition of resonant driving ($\omega = \Omega_1 = \Omega_2$) by setting
\begin{equation}
h_2 = \sqrt{h_{1}^2 + J_{1}^2 - J_{2}^2} \simeq h_1
\end{equation}
with the values of $J_i$ still free as of yet other than being small compared to $h_i$. We enforce the RWA conditions on the driving amplitude of $|\chi_i| \gg
\alpha$ and $||\chi_{1}| - |\chi_{2}|| \gg\alpha$ by taking the integers of Eq.~\eqref{eq:robustness} such that $n_2 = 2n_{1}$ in order to maximize the difference
in Rabi frequencies while keeping both large (which can be ensured via the choice of $n_1$). In the case of detuning-controlled singlet-triplet qubits, due to the
empirically exponential dependence of the exchange interaction on the detuning \cite{Shulman_2012,Dial_2013}, $\delta J_i \propto J_i$ and it is advantageous to choose
small values of $J_i$, but while still maintaining $J_i > j_i$ in order to avoid calling for negative exchange. So, we will choose values of $J_i$ slightly larger than
$j_i$. Without loss of generality, and for the sake of concreteness, we take $j_1 = 2 j_2, J_1 = 2 J_2$. Finally, another physical consideration specific to the
capacitively-coupled singlet-triplet system is the treatment of perturbations in the coupling, $\delta \alpha$. Since $\alpha$ is proportional to the product of the
derivatives of the exchange interactions in each qubit and the proportionality constant is such that $\delta\alpha$ is about two order of magnitude smaller than
$\delta J_i$ \cite{Shulman_2012}, its effects are negligible and can safely be ignored.

We summarize and combine all of the constraints above in the following set of robustness conditions:
\begin{equation}
\begin{gathered}
\label{eq:summary}
h_1, J_1, n_1, \alpha, \theta \;\text{are free and subject to}\\
\alpha \ll j_i < J_i \ll h_i \text{ with } n_1\in \mathbb{Z},\\
\omega = \sqrt{h_1^2+J_1^2}\simeq h_1,\\
J_1 = 2J_2, j_1 =2 j_2,\\
h_2 = \sqrt{h_{1}^2 - 3 J_{1}^2}\simeq h_1\\
m = \text{nint} \left( \frac{\theta}{\pi}\frac{\omega}{\alpha}-\frac{1}{2} \right), \\
t = (m+1/2) \pi/\omega,\\
j_1 =\frac{4 n_{1}\Omega_{1}\omega}{h_{1}(m+1/2)} \simeq \frac{4 n_{1}\omega}{(m+1/2)},\\
\end{gathered}
\end{equation}
where it suffices to meet the approximate equalities due to the condition $J_i \ll h_i$, and $\text{nint}(x)$ is the nearest integer function.
The effect of these constraints on the first-order error channels is shown in Table~\ref{table:TotError2}.  With the parameter choices of Eq.~\eqref{eq:summary}, the
surviving five error channels are left with terms that are approximately proportional to $\frac{\delta j_i}{\alpha}$, $\frac{\delta J_i}{h_i}$,
$\frac{J_i}{h_i}\frac{\delta h_i}{h_i}$, $\frac{J_i}{h_i}\frac{\delta J_i}{h_i}$, $\left(\frac{J_i}{h_i}\right)^2\frac{\delta J_i}{h_i}$, and
$\left(\frac{J_i}{h_i}\right)^2\frac{\delta h_i}{h_i}$. The last four terms in the list are clearly negligible. By invoking the exponential behavior of the exchange
interaction, we have $\delta J_i = \frac{\mathrm{d}J_i}{\mathrm{d}\varepsilon_i}\delta \varepsilon_i \propto J_{i}\delta\varepsilon_i$, which indicates that the second
term in the list is also suppressed for $J_i \ll h_i$. However, the first term in the list is not necessarily small. Errors from $\delta j_i$ accumulate linearly with
the gate time and are, consequently, effectively proportional to $1/\alpha$. Again noting that the empirically exponential nature of the exchange implies $\delta j_i
\propto j_i$, it is possible to avoid unnecessarily large $\delta j_i$ by choosing the free integer $n_1$ that appears in $j_i$ to be as small as possible while still
maintaining the RWA condition of $|\chi_i|\gg\alpha$. The low-frequency content of the remaining $\delta j_i$ error can be removed by inserting a refocusing $\pi$-pulse
about the $x$-axis of each qubit in between two entangling gates. This is a well-known strategy~\cite{Bremner_2005,Hill_2007,Calderon_prl2017}, making use of the fact
that the local $\sigma_{XX}$ insertion commutes with the nonlocal $\sigma_{ZZ}$ phase but anticommutes with the $\sigma_{IZ}$ and $\sigma_{ZI}$ error terms.

Since we are left with only five error channels, extracting the first-order error of the refocused entangling gate like in Eq.~\eqref{eq:nonlocal1} is analytically
straightforward. The refocusing process shuffles these errors among the SU(4) basis elements, some of which appear in the $\sigma_{XX}$, $\sigma_{YY}$, $\sigma_{XY}$,
and $\sigma_{YX}$ channels. These errors commute with the nonlocal $\sigma_{ZZ}$ phase, which suggests that concatenating with a local $\pi$-pulse about the z-axis of
either qubit, e.g. $\sigma_{ZI}$, can be used to further correct the residual errors in the refocused gate.

\subsection{Similar Qubits with $\mathbf{J_i} \gg \mathbf{h_i}$}
\label{subsec:dissimilar}

We follow the same process as before but now we assume that the magnetic field gradients are fixed. Since we are taking $J_i \gg h_i$, the terms in the propagator that
are proportional to $\frac{h_1 h_2}{\Omega_1 \Omega_2}$ are negligibly small. Thus, to generate an entangling gate, it is preferable for us to take the case where
$||\chi_1| - |\chi_2|| \ll \alpha$ (Eq.~\eqref{eq:second_RWA_v1}). Ignoring the negligible terms, the time-evolution is
\begin{align}
\label{eq:lab-sol2}
U(t) &= R_{1}(t) \exp\bigg[-\imath t\frac{\alpha J_1 J_2}{2 \Omega_1 \Omega_2}\left(\sigma_{XX}+\sigma_{YY}\right) \bigg] R_{2}(t)\\
&\simeq R_{1}(t) \exp\bigg[-\imath t\frac{\alpha}{2}\left(\sigma_{XX}+\sigma_{YY}\right) \bigg] R_{2}(t),
\end{align}
where the purely local operators $R_{1}(t)$ and $R_{2}(t)$ are given by
\begin{equation}
\begin{aligned}
R_{1}(t) = &\exp\Bigg[\frac{-\imath}{2}\sum_{i=1}^{2}\phi_{i}\sigma_{Y}^{(i)}\Bigg]
\exp \left[ -\imath \sum_{i=1}^{2} \frac{\omega_i t + \xi_i(t)}{2} \sigma_{X}^{(i)} \right] \\
\times &\exp\left[ -\imath t \sum_{i=1}^{2} \chi_i \sigma_{Z}^{(i)} \right], \\
R_{2}(t) = &\exp\Bigg[\frac{\imath}{2}\sum_{i=1}^{2}\phi_{i}\sigma_{Y}^{(i)}\Bigg].
\end{aligned}
\end{equation}
The error channels for this evolution can be calculated in a similar fashion as in the previous case; the results are reported in  Appendix~\ref{app:errors}.

We proceed to our goal of synchronizing the error terms so that they vanish at the gate time. We can eliminate a number of error terms by choosing our parameters so that
$\sin(\chi_i t)$ and $\cos (\omega_i t + \xi_i(t))$ simultaneously vanish at the gate time. However, as before, a significant amount of error remains in the $\sigma_{IZ}$
and $\sigma_{ZI}$ channels. In this case, though, we cannot simply apply a refocusing $\pi$-pulse since these error channels do not commute with the entanglement generator
$\sigma_{XX}+\sigma_{YY}$. Fortunately, Ref.~\cite{Calderon_prl2017} offers a sequence of 10 local $\pi$-pulses interspersed between short entangling operations that can
deal with these anticommuting errors to first-order while reducing the entanglement generator to $\sigma_{XX}$. Therefore, it is again possible in principle to generate
high-fidelity entangling gates from a combination of stroboscopic decoupling and composite pulses in this parameter regime.

However, we must note that the assumption following Eq.~\eqref{eq:third_rot} of $\delta\Omega_i \ll \chi_i$ is likely unrealistic in this $J_i \gg h_i$ case for the charge
noise levels currently reported in singlet-triplet qubits. Quasistatic fluctuations in the detuning, $\delta\varepsilon$, typically have a standard deviation of several
$\mu$V \cite{Dial_2013} and around $J\sim$GHz this can cause $\delta\Omega_i \sim 10$MHz, whereas in this regime $\chi_i \simeq h_i/4 \sim 10$MHz as well. We estimate that
roughly an order of magnitude decrease in the charge noise strength, down to under a microvolt, would be required in order to safely neglect off-resonance errors. Note that
the previous case of $h_i\gg J_i$ did not have this problem because there $\delta\Omega_i$ is dominated by magnetic noise, which is typically $\sim 10$neV, whereas in that regime
$\chi_i \lesssim j_i/4 \sim 100$neV. Therefore, the case of similar qubits with $h_i \gg J_i$ is a more feasible operating regime for our proposed high-fidelity two-qubit
gates in a double quantum dot singlet-triplet system.  In the context of silicon singlet-triplet qubits with micromagnet gradients, this along with our discussion at the
beginning of Sec.~\ref{sec:Error_Channels} means that the silicon devices must be engineered to either allow enough tunability of the magnetic differences across each qubit
(via dot positioning, etc.) for them to be equalized in situ, or to physically reduce charge noise in the device.  The former seems an easier target.

\section{Simulations}
\label{sec:Simulations}

\begin{table}
\caption{Average \textsc{cphase} fidelity in the presence of 20neV magnetic noise and 8$\mu$V quasistatic charge noise
with a 1/\textit{f}$^{0.7}$ component of 0.9nV/$\sqrt{\text{Hz}}$ at 1MHz.}
\label{table:PulseSequence}
\begin{tabular}{c|c|c}
	\hline
	\hline
	Sequence & $\langle F\rangle_{unoptimized}$ & $\langle F\rangle_{optimized}$\\
	\hline
	No refocusing 			& .768 & .811 \\
	Singly refocused 		& .950 & .974 \\
	Doubly refocused   		& .944 & .996 \\
	\hline
\end{tabular}
\end{table}

We now examine the effects of our optimization in the presence of quasistatic magnetic noise and $1/f^{0.7}$ charge noise \cite{Dial_2013}. We will simulate the fidelity of
\textsc{cphase} gates generated by a single-shot pulse, a single spin echo composite pulse, and a double spin echo composite pulse for both unoptimized and stroboscopically
optimized parameters.

We report in Table~\ref{table:PulseSequence} a summary of the calculated fidelities. The magnetic noise was generated from a normal distribution with a standard deviation of 20neV \cite{Bluhm_2010,Malinowski_2016}. To generate the charge noise, we superimposed
20 random telegraph noises with the appropriate weighting \cite{Kogan_1996} and relaxation times ranging from 1MHz to 1GHz~\cite{Nichol_2017} evenly spaced on a logarithmic
scale with an amplitude of 0.9nV/$\sqrt{\text{Hz}}$ at 1MHz. An additional quasistatic noise component is added to ensure that the integrated power spectral density from
0 to 1MHz is consistent with the experimentally reported noise amplitude of 8$\mu$V~\cite{Dial_2013}. Finally, we translated the noise in detuning $\epsilon$ into noise in
exchange $J$ by using an exponential fit on the data reported in Ref. \cite{Dial_2013}.

We numerically solve for the time-evolution operator using the unapproximated, time-dependent Hamiltonian \eqref{eq:Hamiltonian} with the optimal parameters predicted by the RWA analysis above, and then convert it to a \textsc{cphase} gate by using the perfect local operations prescribed by the RWA, as in the left-hand side of Eq.~\eqref{eq:nonlocal}. Note that for these numerical calculations we do not assume that the RWA is accurate; e.g, we do not assume now that the right-hand side of Eq.~\eqref{eq:nonlocal} holds.  We calculate the average two-qubit gate fidelity \cite{Cabrera_2007}
\begin{equation}
\langle F\rangle = \frac{1}{16}\Bigg[4+\frac{1}{5}\sum_{\sigma_{ij}}\mathrm{tr}[U_{1}\sigma_{ij}
U_{1}^{\dagger} U_{2}\sigma_{ij}U_{2}^{\dagger}]\Bigg],
\end{equation}
where $U_1$ is the ideal \textsc{cphase} and $U_2$ is the actual noisy evolution, which we obtain purely numerically for a given set of parameter values and averaging
over 1000 different noise realizations.  Any error due to the RWA is also included in that fidelity.

\begin{table*}
\caption{Local parameters used in the simulations.}
\label{table:Parameters}
\begin{tabular}{c|c|c|c}
	\hline
	\hline
	Parameters ($\frac{1}{2\pi}$MHz) & Unoptimized, & Optimized, & Optimized, \\
                                     & all cases & no refocusing / singly refocused & doubly refocused\\
	\hline
	$J_1$ & 266 	& 80 		& 150 \\
	$J_2$ & 320 	& 40 		& 75 \\
	$j_1$ & 69 		& 74 		& 147 \\
	$j_2$ & 36 		& 37 		& 73 \\
	$h_1$ & 922 	& 1000 		& 1500 \\
	$h_2$ & 905 	& 1002 		& 1506 \\
	\hline
\end{tabular}
\end{table*}

\begin{figure*}
\centering
\begin{subfigure}
\centering
\includegraphics[scale=.79]{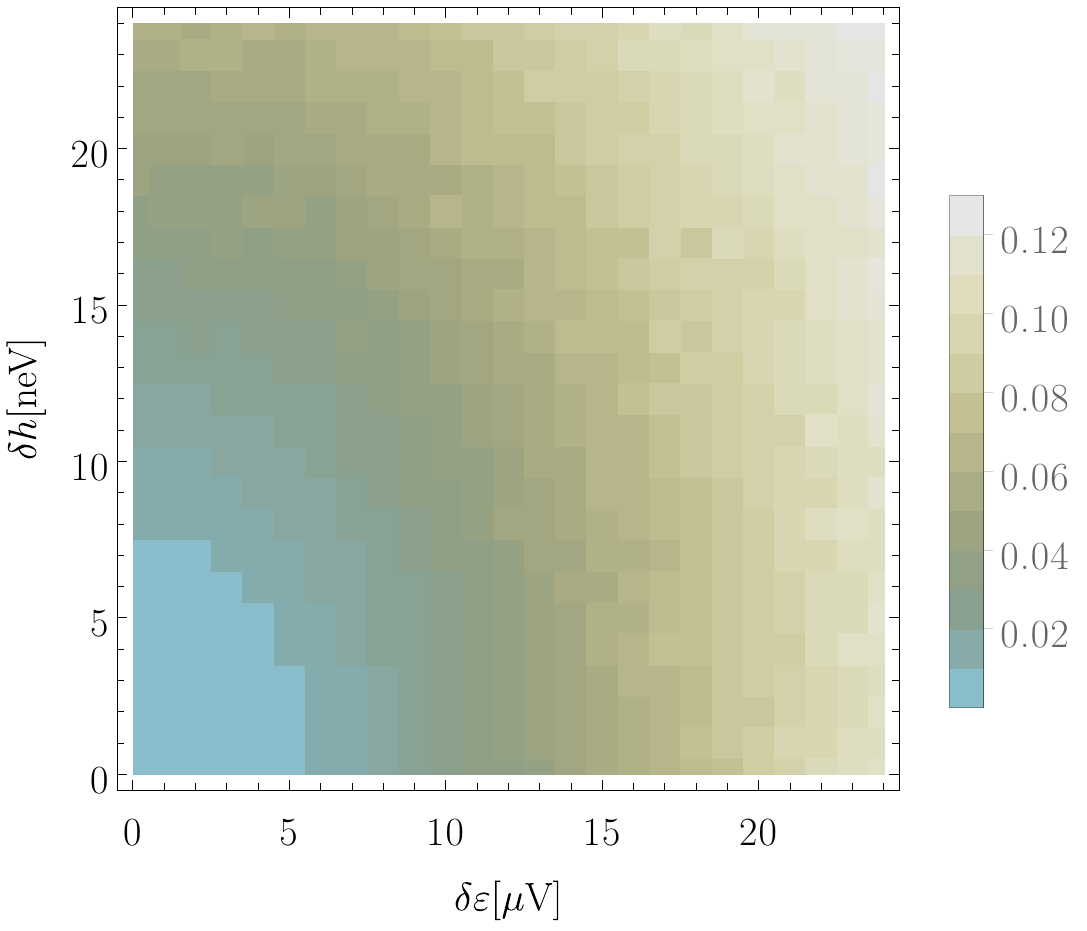}
\end{subfigure}
\begin{subfigure}
\centering
\includegraphics[scale=.79]{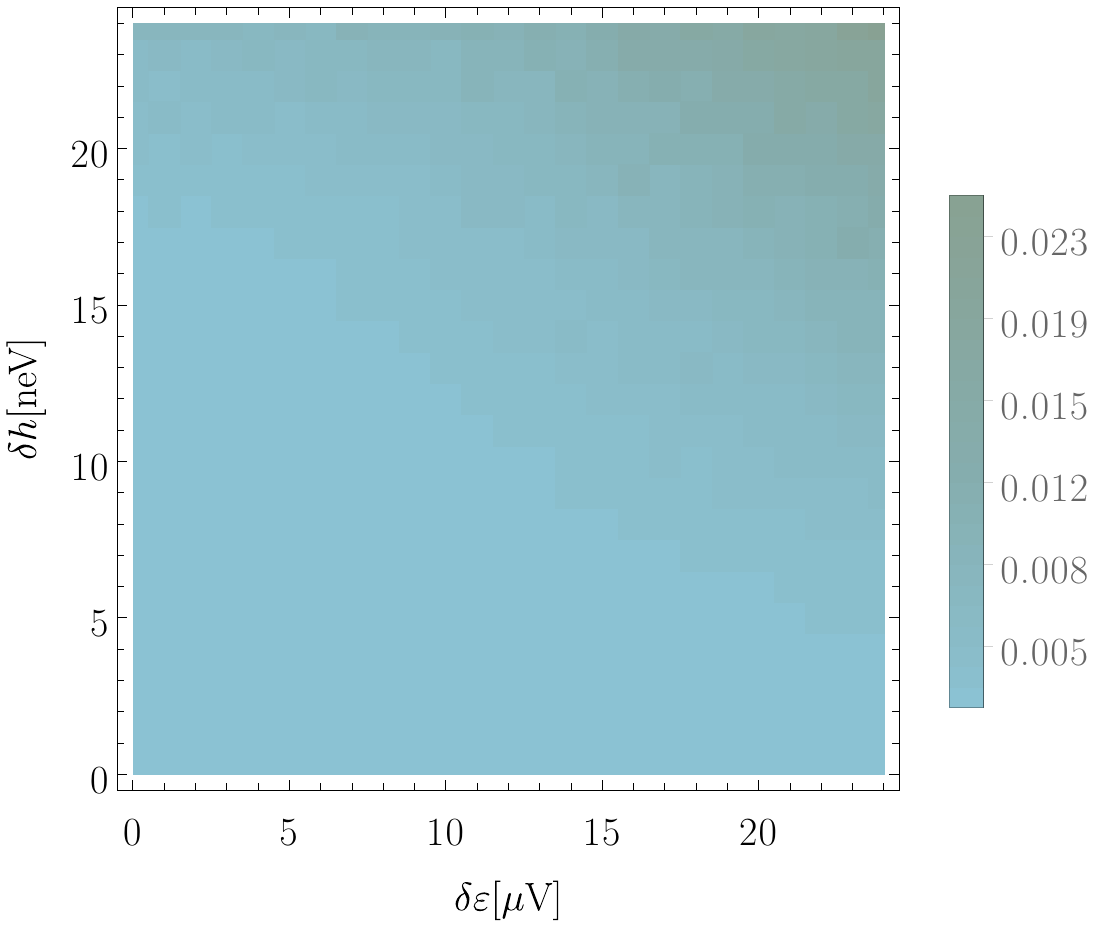}
\end{subfigure}
\caption{Average infidelity as a function of noise strength for the unoptimized (LEFT) and optimized (RIGHT) case after applying a doubly refocusing $\pi$-pulse.
The values in the axes indicate the strength of quasistatic noise. 1/\textit{f} noise is added to the exchange with an amplitude
0.9nV/$\sqrt{\text{Hz}}$ at 1MHz.}
\label{fig:contour}
\end{figure*}

A summary of all the parameter values used in the simulations are provided in Table~\ref{table:Parameters}. We have taken $\alpha = 2\pi\times 2.3$MHz in all cases
for consistency. For all pulse sequences the same unoptimized parameters are used, obtained from Ref. \cite{Calderon_2017} consistent with the range reported in experiment
\cite{Nichol_2017}. On the other hand, the optimized parameters are chosen following the rules in Eq.~\eqref{eq:summary}. We choose the free parameters $h_1 = 1$GHz,
$J_1 = 80$MHz, and $n_1 = 4$ for the no refocusing and singly refocused case, ensuring that $h_1\gg J_1 > j_1$. On the other hand, we take $h_1 = 1.5$GHz, $J_1 = 150$MHz,
and $n_1 = 2$ for the doubly refocused case in order to compensate for the shorter gate time needed. These immediately determine the values of $\omega$, $J_2$, and $h_2$
shown in Table~\ref{table:Parameters}. The value of $\theta$ can either be $\pi/2$, $\pi/4$, or $\pi/8$, depending on which composite pulse sequence is being performed,
as we discuss below.

As previously mentioned, all the simulations target a \textsc{cphase} gate. When applying the singly refocusing pulse, we replace the simple \textsc{cphase}
gate $U_{nl}\left(t_{\pi/2}\right)$ with the composite \textsc{cphase} gate
\begin{equation}
U_{nl}(t_{\pi/4}) \pi_{XX} U_{nl}(t_{\pi/4}) \pi_{XX},
\end{equation}
where $U_{nl}(t_{\theta})$ is the noisy entangling gate targeting a nonlocal phase $\theta$ and $\pi_{ab}$ is a local $\pi$ rotation about the $a$-axis of the first
qubit and the $b$-axis of the second qubit. The doubly refocused composite pulse requires twice as many component gates, but note that the entangling time is not any
longer since each entangling component is shorter,
\begin{equation}
\begin{aligned}
&\left[U_{nl}(t_{\pi/8})  \pi_{XX} U_{nl}(t_{\pi/8}) \pi_{XX}\right] \pi_{ZI} \\
&\qquad \times \left[U_{nl}(t_{\pi/8}) \pi_{XX} U_{nl}(t_{\pi/8}) \pi_{XX}\right] \pi_{ZI} \\
&= U_{nl}(t_{\pi/8}) \pi_{XX} U_{nl}(t_{\pi/8}) \pi_{YX} \\
&\qquad \times U_{nl}(t_{\pi/8}) \pi_{XX} U_{nl}(t_{\pi/8}) \pi_{YX}.
\end{aligned}
\end{equation}

We further examine how our optimization behaves under a range of noise amplitudes. We keep the amplitude of the 1/\textit{f}$^{0.7}$ charge noise component the same as before
for consistency, but we generate quasistatic noise with amplitudes ranging from 0 to 24 neV ($\mu$V) for magnetic (charge) noise. A contour plot of the average infidelity as
a function of quasistatic noise strength for the case of doubly refocused gates is provided in Fig.~\ref{fig:contour}. We find that combining our optimization scheme with the
doubly refocusing pulse yields an order of magnitude improvement in fidelity compared to the unoptimized case. We emphasize that this improvement can be attributed to the
isolation of error onto specific channels presented in Table~\ref{table:TotError2}. In fact, if one can further reduce the average fluctuations in the magnetic field gradient
(e.g. down to 8neV~\cite{Bluhm_2010}), it is possible to generate a \textsc{cphase} gate with average fidelities over 99\% using only the singly refocusing pulse.

\section{Conclusion}
We theoretically analyze the first-order effects of errors in two capacitively-coupled singlet-triplet qubits by perturbing parameters in the time-evolution operator derived
using the RWA. We examined two extreme regions of the parameter space and showed that it is better to operate in the parameter regime where the magnetic field gradient dominates
the exchange than the opposite case.

We find that certain choices of parameter lead to passive, stroboscopic circumvention of errors. This enables the isolation of the errors onto specific basis elements of SU(4),
consequently allowing the application of composite pulse sequence to mitigate the residual errors. Our numerical simulations show that our analytic prescription produces
\textsc{cphase} gates with fidelities above 99\% using only 4 applications of local $\pi$ pulses on each qubit, which is an order of magnitude improvement over an unoptimized
implementation.

This material is based upon work supported by the National Science Foundation under Grant No. 1620740 and by the Army Research Office
(ARO) under Grant No. W911NF-17-1-0287.

\appendix

\section{Effects of Exchange Ramping Evolution}
\label{app:ramp}
In the main text we only considered the case when the exchange $J_{i}(t)$ is controlled using rectangular pulses both in
the beginning and the end of the evolution. Realistically, there is a finite rise time, $\tau$, to go from $J_{i}(-\tau) \approx 0$ up to
$J_{i}(0) = J_i + j_i$, and, since Eq.~\eqref{eq:summary} tells us that the exchange should have gone through a odd number of half cycles at the end of the gate, back down from $J_{i}(t_{gate}) = J_i$ to $J_{i}(t_{gate}+\tau) \approx 0$. We now consider the effects of the evolution during the finite ramp on our optimization scheme.  We will show that the effects are negligible, assuming typical values for the coupling, noise, and rise time.

We choose the well-studied Rosen-Zener pulse shape~\cite{Rosen_1932,Robiscoe_1978,Torosov_2007} for our ramp:
\begin{equation}
J_{i}(t) =
\begin{cases}
J_{i,u} \sech(\frac{2\pi t}{\tau}), &\:\:\: -\tau < t < 0 \\
J_{i,d} \sech(\frac{2\pi (t-t_{gate})}{\tau}), &t_{gate} < t < t_{gate}+\tau,
\end{cases}
\end{equation}
where $J_{i,u} = J_i+j_i$ is the upward ramp amplitude and $J_{i,d} = J_i$ is the downward ramp amplitude. In addition, since there is a rough proportionality between the average capacitive coupling and the average exchanges, $\alpha \propto J_1 J_2$ \cite{Shulman_2012}, the coupling also has a finite ramping time.  However, we take $\tau = 1\,\text{ns}$ which is consistent with
experimental ramp times in spin qubits~\cite{Weperen_2011}, and so a typical coupling that ranges up to $1-2$ MHz \cite{Shulman_2012,Nichol_2017} has a negligible effect on such a short time scale.  Thus the evolution during the ramp is dominated by the local terms, and the ramping Hamiltonian takes the form
\begin{equation}
\label{eq:ramp_H}
\mathcal{H} = \sum_{i=1}^{2} \bigg( \frac{J_{i}(t)}{2}\sigma_{Z}^{(i)} +
\frac{h_{i}}{2}\sigma_{X}^{(i)}\bigg).
\end{equation}

We first consider the case where the exchange is ramped up. We begin by noting that since the spin operators for each qubit commute, then we can separate
the propagator into $U(t) = U_{1}(t)U_{2}(t)$. Each of these propagators are solutions to
\begin{equation}
\label{eq:ramp_su2}
\imath \frac{d}{dt}U_{i}(t) = \bigg(\frac{J_{i,u}\sech\big(\frac{2\pi t}{\tau}\big)}{2}\sigma_{Z}^{(i)} +\frac{h_{i}}{2}\sigma_{X}^{(i)}\bigg) U_{i}(t).
\end{equation}
In order for us to use known analytical results, we first rotate to a frame so that
\begin{equation}
U_{i}(t) = \exp\Big[\imath \frac{\pi}{4}\sigma_{Y}^{(i)}\Big] \exp\Big[\imath t\frac{h_i}{2}\sigma_{Z}^{(i)}\Big] U'_{i}(t).
\end{equation}
This allows us to write two coupled differential equations
\begin{equation}
\begin{gathered}
\begin{cases}
\imath\,\dot{s}(t) = \frac{J_{i,u}\sech\big(2\pi t \big/ \tau\big)}{2} \mathrm{e}^{-\imath h_i t} p(t)\\
\imath\,\dot{p}(t) = \frac{J_{i,u}\sech\big(2\pi t \big/ \tau\big)}{2} \mathrm{e}^{\imath h_i t} s(t),
\end{cases}
\end{gathered}
\end{equation}
where $U'_{i}(t)\psi'(t_o) = (s(t),p(t))^\textsc{t}$ and $\psi'(t_o)$ is the initial wavefunction. Using the results from Refs.~\cite{Rosen_1932,Torosov_2007}, we can write the time-evolution in the rotating frame for $t \leq 0$ as
\begin{equation}
U'_{i}(t) = U_\mathrm{I} \mathbbm{1} + U_\mathrm{X}\sigma_{X}^{(i)} + U_\mathrm{Y}\sigma_{y}^{(i)} + U_\mathrm{Z}\sigma_{Z}^{(i)},
\end{equation}
where
\begin{widetext}
\begin{equation}
\begin{gathered}
U_\mathrm{I} = \frac{1}{2}\Bigg\{{}_{2}F_{1}\bigg[-\frac{J_{i,u} \tau}{2},\frac{J_{i,u} \tau}{2};\frac{1-\imath h_i \tau}{2}; z \bigg] + {}_{2}F_{1}\bigg[-\frac{J_{i,u} \tau}{2},\frac{J_{i,u} \tau}{2};\frac{1+\imath h_i \tau}{2}; z \bigg] \Bigg\}\\
U_\mathrm{X} = \frac{1}{4}J_{i} \tau \sech\bigg[\frac{t}{\tau}\bigg] \left\{
	\frac{\mathrm{e}^{-\imath h_i t}{}_{2}F_{1}\Big[1-\frac{J_{i,u} \tau}{2},1+\frac{J_{i,u} \tau}{2};\frac{3-\imath h_i \tau}{2}; z \Big]}{h_i \tau+\imath}
	-
	\frac{\mathrm{e}^{\imath h_i t}{}_{2}F_{1}\Big[1-\frac{J_{i,u} \tau}{2},1+\frac{J_{i,u} \tau}{2};\frac{3+\imath h_i \tau}{2}; z \Big]}{h_i \tau-\imath}\right\}\\
U_\mathrm{Y} = \imath\frac{1}{4}J_{i} \tau \sech\bigg[\frac{t}{\tau}\bigg] \left\{
	\frac{\mathrm{e}^{-\imath h_i t}{}_{2}F_{1}\Big[1-\frac{J_{i,u} \tau}{2},1+\frac{J_{i,u} \tau}{2};\frac{3-\imath h_i \tau}{2}; z \Big]}{h_i \tau+\imath}
	+
	\frac{\mathrm{e}^{\imath h_i t}{}_{2}F_{1}\Big[1-\frac{J_{i,u} \tau}{2},1+\frac{J_{i,u} \tau}{2};\frac{3+\imath h_i \tau}{2}; z \Big]}{h_i \tau-\imath}\right\}\\
U_\mathrm{Z} = \frac{1}{2}\Bigg\{{}_{2}F_{1}\bigg[-\frac{J_{i,u} \tau}{2},\frac{J_{i,u} \tau}{2};\frac{1+\imath h_i \tau}{2}; z \bigg] - {}_{2}F_{1}\bigg[-\frac{J_{i,u} \tau}{2},\frac{J_{i,u} \tau}{2};\frac{1-\imath h_i \tau}{2}; z \bigg] \Bigg\},
\end{gathered}
\end{equation}
\end{widetext}
and ${}_{2}F_{1}[a,b;c;d]$ is Gauss's hypergeometric function and $z = \frac{1}{2}\left(1+\tanh\left[\frac{t}{\tau}\right]\right)$. We note that $U'_{i}(t)$ satisfies the initial condition $U'_{i}(-\infty) = \mathbbm{1}$. In order to get the actual solution to equation~\eqref{eq:ramp_su2} with $U_{i}(-\tau) = \mathbbm{1}$, we use the composition
property of time-evolution operators:
\begin{equation}
U_{i}(t;-\tau) = U_{i}(t;-\infty) U_{i}^{\dagger}(-\tau;-\infty),
\end{equation}
where $U(t;t_o)$ indicates the evolution from $t_o$ to $t$. More explicitly, the upward ramp propagator that corresponds to the Hamiltonian in
equation~\eqref{eq:ramp_H} is approximately given by
\begin{equation}
\begin{aligned}
U_{u}(t;-\tau) &= \exp\left[\imath \frac{\pi}{4}\sum_{i=1}^{2}\sigma_{Y}^{(i)}\right] \exp\left[\imath t \sum_{i=1}^{2}h_{i}\sigma_{Z}^{(i)}\right] U'_{1}(t)\\
&\times U'^{\dagger}_{1}(-\tau) U'_{2}(t) U'^{\dagger}_{2}(-\tau)\\
&\times \exp\left[\imath \tau \sum_{i=1}^{2}h_{i}\sigma_{Z}^{(i)}\right] \exp\left[-\imath \frac{\pi}{4}\sum_{i=1}^{2}\sigma_{Y}^{(i)}\right].
\end{aligned}
\end{equation}

To solve for the downward ramp evolution, we first note that there is a relationship between the upward and
downward ramp Hamiltonian when their amplitudes are similar: $\mathcal{H}_{d}(t) = \mathcal{H}_{u}(t_{gate}-t)$.
Using this, then we can write the time-evolution of the downward ramp as
\begin{equation}
U_{d}(t) = \mathcal{T} \exp\left[ -\imath \int_{t_{gate}}^{t}H_{d}(t')\mathrm{d}t' \right].
\end{equation}
where $\mathcal{T}$ denotes the time-ordering operator. Using a simple change of variable and using the composition property of time-evolution operators, we
can express the evolution of the downward ramp in terms of the upward ramp:
\begin{align*}
U_{d}(t) &= \mathcal{T} \exp\left[ -\imath \int_{t_{gate}}^{t}H_{u}(t_{gate}-t')\mathrm{d}t' \right]\\
		 &= \mathcal{T} \exp\left[ \imath \int_{0}^{t_{gate}-t}H_{u}(t'')\mathrm{d}t'' \right]\\
		 &= \mathcal{T} \exp\left[ \imath \int_{-\tau}^{t_{gate}-t}H_{u}(t'')\mathrm{d}t'' \right]\\
		 &\quad \times \mathcal{T} \exp\left[ \imath \int_{0}^{-\tau}H_{u}(t'')\mathrm{d}t'' \right]\\
		 &= \mathcal{T} \exp\left[ \imath \int_{-\tau}^{t_{gate}-t}H_{u}(t'')\mathrm{d}t'' \right]\\
		 &\quad \times \left(\mathcal{T} \exp\left[ \imath \int_{-\tau}^{0}H_{u}(t'')\mathrm{d}t'' \right]\right)^\dagger\\
		 &= \bar{U}_{u}(t_{gate}-t;-\tau) \bar{U}_{u}^{\dagger}(0,-\tau).
\end{align*}
where the bar indicates change from $J_{i,u} \rightarrow -J_{i,d}$, and $h_i \rightarrow -h_i$. Therefore, the downward ramp propagator is given by
\begin{equation}
\begin{aligned}
U_{d}(t;t_{gate}) &= \exp\left[\imath \frac{\pi}{4}\sum_{i=1}^{2}\sigma_{Y}^{(i)}\right] \exp\left[-\imath (\tau-t) \sum_{i=1}^{2}h_{i}\sigma_{Z}^{(i)}\right]\\
&\times \bar{U}'_{1}(\tau-t)\bar{U}'_{1}(0)\bar{U}'_{2}(\tau-t)\bar{U}'_{2}(0) \\
&\times \exp\left[-\imath \frac{\pi}{4}\sum_{i=1}^{2}\sigma_{Y}^{(i)}\right].
\end{aligned}
\end{equation}

Now that we have an analytical expression for the ramp propagators, we can finally address how they affect the error channels and our optimization. In the presence of noise,
it can be verified numerically with the parameters provided in section~\ref{sec:Simulations} that perturbations in $J_i$ results in infidelities that are one to two orders of magnitude smaller than the infidelities we report in the main text. This can be mainly attributed to the fact that $h_i \gg J_i$ and $\tau$ is relatively short. Thus, the dominant source of error in the ramp evolution is due to perturbations in the magnetic gradient $\delta h_i$. However, if we assume $1$ ns ramp times and a standard deviation $\delta h_i = 8$neV~\cite{Bluhm_2010}, the resulting infidelities are also found to
be an order of magnitude smaller than those discussed in the main text. Thus, provided that $\delta h_i \tau$ is much less than the remaining errors in Table~\ref{table:TotError2}, then the errors associated with the ramp can be neglected.

Finally, we address how the unperturbed ramp evolution affect the error channels. The total evolution of the qubits is given by
\begin{equation}
U(t) = U_{u}(t) R_{1}(t) \exp\left[-\imath t \frac{\alpha h_1 h_2}{\Omega_1 \Omega_2}\sigma_{ZZ}\right] R_{2}(t) U_{d}(t).
\end{equation}
We can rewrite this into
\begin{equation}
\begin{aligned}
U(t) &= U_{u}(t) \big( R_{1}(t) R_{1}^{\dagger}(t) \big) R_{1}(t) \exp\left[-\imath t \frac{\alpha h_1 h_2}{\Omega_1 \Omega_2}\sigma_{ZZ}\right] \\
&\times R_{2}(t) \big( R_{2}^{\dagger}(t) R_{2}(t) \big) U_{d}(t).
\end{aligned}
\end{equation}
We can further rewrite this in terms of our optimized gate given in equation~\eqref{eq:nonlocal1}:
\begin{equation}
U(t) = U_{u}(t) R_{1}(t) U_{nl}(t) R_{2}(t) U_{d}.
\end{equation}
Since $U_{u}$ and $U_{d}$ are purely local operations and provided that the ramp errors are negligible, then applying an initial local rotation
$R_{1}^{\dagger}(t) U_{u}^{\dagger}(t)$ and a final local rotation $U_{d}^{\dagger}(t) R_{2}^{\dagger}(t)$ ensures that our optimized gate
$U_{nl}(t)$ and its errors are unperturbed by the ramps.

\section{Error Channels}
\label{app:errors}
We present here a table of error channels for the dissimilar qubit case in Section~\ref{sec:Error_Channels}.
\begin{table*}
\caption{First-order errors for the similar qubit case with $J_i \gg h_i$. Due to the complexity of the error channels,
we had only shown the errors due to fluctuations in the first qubit. To find the effects of perturbations in the second qubit, one need only
generate a second table where the labels are swapped (1 $\leftrightarrow$ 2 and $\sigma_{ij} \leftrightarrow \sigma_{ji}$).}
\label{table:TotErrorDiss}
\begin{center}
\begin{tabular}{c c}
	\hline
	\hline
	$\sigma_{IX}$ & $0$\\
	$\sigma_{IY}$ & $0$\\
	$\sigma_{IZ}$ & $\left( \frac{\imath \left( (J_1 \delta h_1 - h_1 \delta J_1)\sin(\omega_1 t+\xi_1) - 2 \Omega_1^2 t
	\big(\frac{\partial \chi_1}{\partial h_1}\delta h_1 + \frac{\partial \chi_1}{\partial J_1}\delta J_1\big) \right)}{2\Omega_1^2} - \imath t \frac{\partial \chi_1}{\partial j_1} \delta j_1 \right)
	\sin^{2}\left( \frac{J_1 J_2 \alpha t}{\Omega_1 \Omega_2}\right)$\\
	$\sigma_{XI}$ & $\left( \left( \frac{\imath(h_1 \delta J_1 - J_1\delta h_1) \cos(\omega_1 t + \xi_1)}{2\Omega_{1}^{2}}
	- \frac{\imath(h_1 \delta h_1 + J_1 \delta J_1)}
	{4 \chi_1 \Omega_1}\right) \sin(2\chi_1 t) - \frac{\imath}{2}\cos(2\chi_1 t) \left(\frac{\partial \xi_1}{\partial h_1}\delta h_1 + \frac{\partial \xi_1}{\partial J_1}\delta J_1
	+ \frac{\partial \xi_1}{\partial j_1}\delta j_1\right)\right) \cos\left(\frac{J_1 J_2 \alpha t}{\Omega_1 \Omega_2}\right)$\\
	$\sigma_{XX}$ & $\frac{\imath h_1 J_2 \alpha t (J_1 \delta h_1 - h_1 \delta J_1)}{2\Omega_{1}^{3}\Omega_{2}}
	-\imath \frac{J_1 J_2 t \delta\alpha}{2\Omega_1 \Omega_2}$\\
	$\sigma_{XY}$ & $\left( \frac{\imath \left( (h_1 \delta J_1 - J_1 \delta h_1)\sin(\omega_1 t+\xi_1) + 2 \Omega_1^2 t
	\big(\frac{\partial \chi_1}{\partial h_1}\delta h_1 + \frac{\partial \chi_1}{\partial J_1}\delta J_1\big) \right)}{4\Omega_1^2}
	+\frac{\imath}{2} t \frac{\partial \chi_1}{\partial j_1}\delta j_1\right) \sin\left( \frac{2J_1 J_2 \alpha t}{\Omega_1 \Omega_2}\right)$\\
	$\sigma_{XZ}$ & $0$\\
	$\sigma_{YI}$ & $\left( \frac{\imath(h_1 \delta J_1 - J_1 \delta h_1)\left( \cos(2\chi_1 t)\cos(\omega_1 t+\xi_1)\right)}{2\Omega_1^2}
	+ \frac{\imath(h_1 \delta h_1 + J_1 \delta J_1)\sin^{2}(\chi_1 t)}
	{2 \chi_1 \Omega_1} + \frac{\imath}{2}\sin(2\chi_1 t) \left(\frac{\partial \xi_1}{\partial h_1}\delta h_1
	+ \frac{\partial \xi_1}{\partial J_1}\delta J_1 + \frac{\partial \xi_1}{\partial j_1}\delta j_1 \right)\right) \cos\left(\frac{J_1 J_2 \alpha t}{\Omega_1 \Omega_2}\right)$\\
	$\sigma_{YX}$ & $\left( \frac{\imath \left( (J_1 \delta h_1 - h_1 \delta J_1)\sin(\omega_1 t+\xi_1) - 2 \Omega_1^2 t
	\big(\frac{\partial \chi_1}{\partial h_1}\delta h_1 + \frac{\partial \chi_1}{\partial J_1}\delta J_1\big) \right)}{4\Omega_1^2}
	 -\frac{\imath}{2} t \frac{\partial \chi_1}{\partial j_1}\delta j_1\right) \sin\left(\frac{2J_1 J_2 \alpha t}{\Omega_1 \Omega_2}\right)$\\
	$\sigma_{YY}$ & $\frac{\imath h_1 J_2 \alpha t (J_1 \delta h_1 - h_1 \delta J_1)}{2\Omega_{1}^{3}\Omega_{2}}
	-\imath \frac{J_1 J_2 t \delta\alpha}{2\Omega_1 \Omega_2}$\\
	$\sigma_{YZ}$ & $0$\\
	$\sigma_{ZI}$ & $\left(\frac{\imath \left( (J_1 \delta h_1 - h_1 \delta J_1)\sin(\omega_1 t+\xi_1) - 2 \Omega_1^2 t
	\big(\frac{\partial \chi_1}{\partial h_1}\delta h_1 + \frac{\partial \chi_1}{\partial J_1}\delta J_1\big) \right)}{2\Omega_1^2}
	 - \imath t \frac{\partial \chi_1}{\partial j_1}\delta j_1\right)\cos^{2}\left( \frac{J_1 J_2 \alpha t}{\Omega_1 \Omega_2}\right)$\\
	$\sigma_{ZX}$ & $\left(\frac{\imath(J_1 \delta h_1 - h_1 \delta J_1)\cos(2\chi_1 t)\cos(\omega_1 t + \xi_1)}{2\Omega_1^2}
	- \frac{\imath(h_1 \delta h_1 + J_1 \delta J_1)\sin^{2}(\chi_1 t)}
	{2 \chi_1 \Omega_1} - \frac{\imath}{2}\sin(2\chi_1 t) \left(\frac{\partial \xi_1}{\partial h_1}\delta h_1
	+ \frac{\partial \xi_1}{\partial J_1}\delta J_1 + \frac{\partial \xi_1}{\partial j_1}\delta j_1\right)\right) \sin\left(\frac{J_1 J_2 \alpha t}{\Omega_1 \Omega_2}\right)$\\
	$\sigma_{ZY}$ & $\left(\left( \frac{\imath(h_1 \delta J_1 - J_1 \delta h_1)\cos(\omega_1 t + \xi_1)}{2\Omega_1^2}
	- \frac{\imath(h_1 \delta h_1 + J_1 \delta J_1)}{4 \chi_1 \Omega_1}\right) \sin(2\chi_1 t)
	- \frac{\imath}{2}\cos(2\chi_1 t) \left(\frac{\partial \xi_1}{\partial h_1}\delta h_1 + \frac{\partial \xi_1}{\partial J_1}\delta J_1
	+ \frac{\partial \xi_1}{\partial j_1}\delta j_1\right)\right) \sin\left(\frac{J_1 J_2 \alpha t}{\Omega_1 \Omega_2}\right)$\\
	$\sigma_{ZZ}$ & $0$\\
	\hline
\end{tabular}
\end{center}
\end{table*}

\bibliography{stroboscopic_robust_gates_ref}
\bibliographystyle{apsrev4-1}

\end{document}